# Tuning Edge States in Strained-Layer InAs/GaInSb Quantum Spin Hall Insulators

Lingjie Du[1†], Tingxin Li[1,2†], Wenkai Lou[3], Xingjun Wu[2], Xiaoxue Liu[2], Zhongdong Han[2], Chi Zhang[2,4], Gerard Sullivan[5], Amal Ikhlassi[5], Kai Chang[3], Rui-Rui Du[1,2,4*]

[1]*Department of Physics and Astronomy, Rice University, Houston, Texas 77251-1892, USA*
[2]*International Center for Quantum Materials, School of Physics, Peking University, Beijing 100871, China*
[3]*SKLSM, Institute of Semiconductors, Chinese Academy of Sciences, Beijing 100083, China*
[4]*Collaborative Innovation Center of Quantum Matter, Beijing 100871, China*
[5]*Teledyne Scientific and Imaging, Thousand Oaks, California 91603, USA*

† These authors contributed equally to this work.
* Corresponding author: Rui-Rui Du at rrd@rice.edu

*Abstract*

We report on a class of quantum spin Hall insulators (QSHIs) in strained-layer InAs/GaInSb quantum wells, in which the bulk gaps are enhanced by up to five folds as compared to the binary InAs/GaSb QSHI. Remarkably, with consequently increasing edge velocity, the edge conductance at zero and applied magnetic fields manifests time reversal symmetry (TRS) - protected properties consistent with $Z_2$ topological insulator. The InAs/GaInSb bilayers offer a much sought-after platform for future studies and applications of the QSHI.

*Introduction* TRS protected quantum spin Hall effect (QSHE) is predicted in a two-dimensional topological insulator [1-4] with a topological number $Z_2$. The transport evidence for QSHE was first observed [5] in HgTe/CdTe quantum well (QW) with its edge conductance quantized to the theoretical value. To date the leading materials systems are made of semiconductor QWs, *i.e.*, HgTe/CdTe QW and inverted InAs/GaSb QWs; both are described by the Bernevig-Hughes-Zhang model [4]. In InAs/GaSb QWs, wave-function hybridization between InAs and GaSb layers dominates the bulk and opens a minigap $\Delta$ [6], while a Kramer's pair of spin-momentum-locked edge states emerges on the device perimeters [7]. Relevant experiments are reported in refs [8-17]. The charge transport in helical edges is dissipationless, owning to the notion that the helical property prevents charge backscattering. On the other hand, theories [18-20] taking into account electron-electron interactions and correlations suggest that certain many-body scattering processes may exist, which should lead to dissipation.

In the inverted InAs/GaSb bilayer system, the ground electron sub-band in InAs well and the ground hole sub-band in GaSb well cross at certain wave-vectors $k_{cross}$. Spatially separated electrons and holes are strongly coupled at this crossing point due to the tunneling between the two wells; consequently, a hybridization gap $\Delta$ is opened at $k_{cross}$, which is the bulk gap of the QSHI. The density of the charge neutral point (CNP) in the inverted regime is referred to as $n_{cross} = k_{cross}^2/2\pi$. The degree of band inversion can be tuned by QW widths and gate voltages [6, 8-16], and it has dramatic influences on the bulk transport properties. In the deeply inverted regime where typically $n_{cross}$ above $\sim 2 \times 10^{11}$ cm$^{-2}$, there always exist considerable residual states in the hybridization gap thus the bulk of InAs/GaSb QWs is not truly insulating [8,9,12,13], which limits the studies and applications of QSHE.

On the other hand, in the shallowly inverted regime ($n_{cross}$ below $\sim 1 \times 10^{11}$ cm$^{-2}$), the bulk is insulating to a high degree and quantized helical edge conductance plateaus were observed [10,15]. Surprisingly, the quantized conductance plateaus persist under external magnetic fields, in contrast with the theoretical expectations for TRS protected helical edge states [10]. On a general ground, Coulomb interactions of electron-hole pairs dominate over hybridization effects in such a dilute limit [21], leading to the possibility of a 2D excitonic ground state [21-23]. Moreover, here the edge Fermi velocity $v_F \sim \Delta/2\hbar k_{cross}$ is unusually small, in the range of $\sim 2 \times 10^4$ ms$^{-1}$ to $\sim 5 \times 10^4$ ms$^{-1}$, indicating that the edge states are in a strongly interacting regime [18-20,24]. Overall, while the quantized edge transport has been observed in micrometer size samples of shallowly inverted InAs/GaSb, the resilience to external magnetic field and the observed length dependence in long samples are not account for by single-particle model. From an experimental perspective, it is much desirable to develop a plain vanilla QSHI with properties dominated by single-particle physics. Ideally, to some degree the interaction effects may be set in by tuning experimental parameters such as $v_F$.

In this Letter, we report on a QSHI in strained-layer InAs/GaInSb QWs, which clearly manifests TRS protected properties. Due to the band structural changes from strain effect, QWs can be made narrower, leading to stronger overlaps between electron and hole wave functions. This effect results in insulating hybridization gaps at low temperatures even when the $n_{\text{cross}}$ is larger than $3 \times 10^{11}$ cm$^{-2}$. In addition, the helical edge conductance decreases under either perpendicular or in-plane magnetic fields, indicating the opening of mass gaps in the edge states. Remarkably, we found that the edge conductance and the magnetic response are correlated with $v_F$, which could be well controlled by lattice strain and the gate voltages.

***Strain effect in InAs/GaInSb*** Strain-engineering is a common way to modify the band structure and physical properties for semiconductor materials, and recently for topological materials [25,26]. Specific to InAs/GaInSb system, strained-layer InAs/Ga$_{1-x}$In$_x$Sb superlattice (SL) infrared detectors were proposed [27] by Smith and Maihiot in 1987. By alloying GaSb (lattice constant about 6.1 Å) with InSb (6.4 Å), because of the strain in the growth plane, the energy of the conduction band (CB) in InAs shift downward while the energy level of valence band (VB) in GaInSb splits into heavy hole (HH) level and light hole (LL) level, respectively, where the energy of the HH level is higher than the original top VB in GaSb. As a result, to reach a fixed energy band gap, the layers of InAs/GaInSb SL are made narrower than InAs/GaSb SL thereby increasing the optical absorption efficiency. Such strain-engineering has led to the invention of high-performance long-wave length SL infrared detectors [28].

Similar physics idea may guide the construction of a large-gap QSHI. Based on the strain effects described above, we can reach the same inverted band structure with narrower QWs in strained-layer InAs/GaInSb, comparing to unstrained InAs/GaSb. The hybridization-induced gap should increase in such narrower QWs primarily due to the enhanced overlap of electron and hole wavefunctions. In addition, due to the energy splitting of the HH and LH in GaInSb, the Fermi surface of electrons would better match with the Fermi surface of holes, which also help to reduce the residual non-hybridized carriers.

Fig. 1(a)-1(c) shows calculated band structure of strained InAs/Ga$_{1-x}$In$_x$Sb QWs with different indium concentrations (x = 0.20, 0.25, 0.32) by 8-band Kane model. The results indicate that a ~ 20 meV hybridization gap could be achieved in the [100] direction in InAs/Ga$_{0.68}$In$_{0.32}$Sb QWs, which is about five-fold enhancement from the value ~ 4 meV in unstrained InAs/GaSb QWs. Depending on gating conditions, measured bulk gap is around this value. The wafers we used for the present experiment was grown by molecular beam epitaxy (MBE). As an example, the structure of a 9.5 nm InAs/ 4 nm $Ga_{0.75}In_{0.25}Sb$ QWs is shown in Fig. 1(e). Fig. 1(f) is a transmission electron microscope (TEM) photograph of an $InAs/Ga_{0.68}In_{0.32}Sb$ wafer; it shows that the crystalline structure remains coherent across the heterostructure interfaces regardless of ~ 1.5 % in-plane strain.

*Transport properties of bulk states in strained-layer InAs/GaInSb QWs* In order to directly measure the bulk conductance, we fabricate dual-gated Corbino devices. In this case, the edge conductance is shunted and has no contribution to the signals. Fig. 2(a)&2(c), and 2(b)&2(d) shows the traces of the conductivity versus front-gate voltage $V_{front}$ measured from a Corbino device made by the $InAs/Ga_{0.75}In_{0.25}Sb$ QWs at temperature $T \sim 20$ mK, with back gate voltage $V_{back} = 0$ V and $V_{back} = 4$ V, respectively. At the CNP, the conductivity show dips, indicating the entrance into an energy gap. For more positive $V_{back}$, the bulk band becomes more inverted, resulting in a less insulating bulk. Nevertheless, the bulk conductivity is still negligible at low temperature, about 100 MΩ per square at 20 mK for the $V_{back} = 0$ V case, and about 25 MΩ per square at 20 mK for the $V_{back} = 4$ V case. Note that even for the $V_{back} = 0$ V case, the $n_{cross}$ value of this wafer is larger than $2 \times 10^{11}$ cm$^{-2}$, corresponding to the modestly deep-inverted regime. Hybridization gaps with residual conductivity have been commonly reported in deeply inverted InAs/GaSb QWs [6,8,12,13]; this is the first time that a substantially insulating hybridization gap is observed at low temperature.

Electron-hole hybridization are most favored when the Fermi momentum of electrons $k^e_F$ and holes $k^h_F$ are equal. Under in-plane magnetic field $B_{//}$, applied along $x$ axis of the example, Lorenz force gives tunneling carriers additional momentum along $y$ axis, resulting in a relative shift of band dispersions $\Delta k_y = -eB\Delta<z>/h$, (tunneling distance $\Delta<z>$ is limited by one-half thickness of the QWs). Consequently, carrier hybridization is suppressed due to momentum-mismatch, rendering the QWs as a bilayer-semimetal. As shown in Fig. 2(a) and 2(b), the gap at CNP is gradually closed with an increasing $B_{//}$. Similar behaviors have also been observed in the $InAs/Ga_{0.80}In_{0.20}Sb$ QWs (Fig 2(e)) and $InAs/Ga_{0.68}In_{0.32}Sb$ QWs (Fig 2(f)). This observation agrees with the behavior of a hybridization gap under in-plane magnetic field, but in contrast to the behavior of the insulating gap observed in a shallowly-inverted InAs/GaSb QW [23], where the bulk gap does not show sign of closing in a very high field. Under perpendicular magnetic field $B_\perp$, the bulk becomes more insulating due to localization effects, as shown in Fig. 2(c) and 2(d).

Information of the bulk gaps can be further obtained from temperature dependent conductance. Fig. 2(g) shows the Arrhenius plots of Corbino devices made of strained-layer InAs/Ga$_{1-x}$In$_x$Sb QWs (x = 0.20, 0.25, and 0.32) and the shallowly inverted InAs/GaSb QWs (data adapted from Ref. 10 at $B = 0$ T). It lacks the exponential dependences in the tail regime for the strained layer wafers; the transport there is more like variable-range hopping. Indeed, this is a characteristic feature for transport in hybridization gap, as discussed in ref. 8. At higher temperatures, the hybridization gap values can be estimated by fitting the Arrhenius plots, which is ~ 66 K for the shallowly inverted InAs/GaSb QWs, ~ 120 K for the InAs/Ga$_{0.80}$In$_{0.20}$Sb QWs, ~ 130 K for the InAs/Ga$_{0.75}$In$_{0.25}$Sb QWs, and ~ 250 K for the InAs/Ga$_{0.68}$In$_{0.32}$Sb QWs. Overall, larger hybridization gaps have been achieved by strain-engineering, in reasonable agreement with the calculations.

***Controllable helical edge states with long characteristic length*** We now turn to the helical edge properties of strained-layer InAs/Ga$_{0.75}$In$_{0.25}$Sb QWs. Fig. 3(a) shows the longitudinal resistance $R_{xx}$-$V_{\text{front}}$ traces of a 100 μm × 50 μm Hall bar device with various $V_{\text{back}}$ at $T \sim 20$ mK. Here the measured $R_{xx}$ is solely resulted from the edge channels, since the bulk is fully insulating at such low $T$. At $V_{\text{back}} = 0$ V, the resistance peak is about 115 kΩ, corresponding to a characteristic length $\lambda_\varphi$ (refers to a length scale at which dissipationless edge transport breaks down and counter propagating spin-up and spin-down channels equilibrate) about 11 μm. The $\lambda_\varphi$ of different devices made by this wafer typically range from ~ 5 μm to 10 μm, significantly longer than those in previous studies [5,10] of the QSHI. Remarkably, the $\lambda_\varphi$ can be tuned by gate: as shown in Fig. 3(a) the resistance peak values gradually decreases with decreasing $V_{\text{back}}$ (namely, less inverted), indicating that the $\lambda_\varphi$ increase from ~ 6 μm at $V_{\text{back}} = 4$ V to ~ 11 μm at $V_{\text{back}} = 0$ V. (Note for this device the backgate bias was limited within 4V and 0V). The insets of Fig. 3(a) illustrate the $\lambda_\varphi$ and the $n_{\text{cross}}$ (deduced from magneto-transport data) versus $V_{\text{back}}$.

According to the definition of characteristic length, if the device edge length is shorter than $\lambda_\varphi$, the edge, conductance measured should be quantized to $2e^2/h$. This is indeed confirmed in a Hall bar device of length 10 μm, as shown in Fig. 3(b). As the $\lambda_\varphi$ is being tuned from 6 μm to 11 μm, the $R_{xx}$ decreases, and finally reaches a plateau of 12.9 kΩ with a reasonable accuracy.

A plausible explanation for above data is related to the interaction effects [18-20] in the helical edge state. At more positive $V_{\text{back}}$, the bulk band becomes more inverted hence a larger $k_{\text{cross}}$ and a roughly constant $\Delta$ [7]; overall this would lead to a smaller Fermi velocity $v_F \sim \Delta/2\hbar k_{\text{cross}}$ of the helical edge states, resulting in more prominent interaction effects for the edge states. The backscattering processes would enhance when the electron-electron interactions become stronger, thus the helical edge states exhibit a shorter characteristic length in the more inverted case.

***TRS protected helical edge states*** In general, applying magnetic field will break the TRS and open a gap in the helical edge states. 1D massless Dirac fermion could be tuned into 1D massive fermion allowing for backscattering, thereby the helical edge resistance will increase. However, in previous studies [10] of shallowly-inverted InAs/GaSb QWs, the quantized conductance plateaus are found to persist under external magnetic fields, in contrast with the theoretical expectations for TRS protected helical edge states.

Remarkably, for all devices made by strained-layer InAs/GaInSb QWs, the helical edge conductance show clear magnetic field dependence. Specifically, for a 3 μm × 1.5 μm Hall bar device made by the InAs/Ga$_{0.75}$In$_{0.25}$Sb QWs, a quantized plateau of $h/2e^2$ has been observed at zero magnetic field, as shown in Fig. 4(a) and 4(b). Under a perpendicular magnetic field, as shown in Fig. 4(a), the plateau values ($R_{CNP}$) increase at first ($B_\perp$ below 5 T) due to TRS breaking, then decrease at higher $B_\perp$, indicating the edge states undergoing a transition from helical edge states to chiral edge states [10]. Similar behaviors were observed for the longer sample of 100 μm × 50 μm Hall bar, as shown in Fig. 4(c) and 4(d).

The response to an in-plane field $B_{//}$ shows an interesting behavior. Under a small $B_{//}$ up to 3 T (Fig. 4(b)), the measured resistance of the 3 μm × 1.5 μm Hall bar increases due to TRS breaking. For $B_{//}$ above 3 T, we observed that the sample resistance decreases with $B_{//}$, primarily because the bulk becomes conductive under higher $B_{//}$ (see Fig. S4 and S5 of Supplemental Material [29]).

Following comments are in order. 1) The helical edge states here should be described as a weakly interacting 1D helical liquid without axial spin symmetry, *i.e.*, spin S$_z$ is momentum dependent [31]. Consequently, additional TRS-allowed inelastic scattering channels exist. Our data show, that an external magnetic field (either $B_\perp$ or $B_{//}$, or in a combination of both) would cause the edge resistance $R_{xx}$ to increase, qualitatively consistent with this spin texture picture; 2) Maciejko *et al* [32] studied the combined effect of disorder and TRS breaking on QSH edge transport. They conclude that in the absence of TRS, the edge liquid is topologically equivalent to a spinless 1D quantum wire, and therefore subject to Anderson localization by disorder. We note that (refer to Fig. 4) under magnetic field the $R_{xx}$ increases throughout the bulk gap, indicating bulk disorder may play a role in localization of the edge states [32]; and 3) It appears that the response of $R_{xx}$ to a magnetic field correlates with $\lambda_\varphi$ (hence with $v_F$). This can be seen in a 100 μm × 50 μm Hall bar under $B_\perp = 1$ T at $V_{back} = 0$ V (Fig. 4(c)) and $V_{back} = 4$ V (Fig. 4(d)), where $R_{xx}$ increases by 41% (32%) for $\lambda_\varphi \sim 10.7$ μm (6.8 μm), respectively.

***Discussion on Luttinger parameter K***   One of the most attractive features of the strained-layer InAs/GaInSb system is the relatively large hybridization gap, and the gap size can be well controlled by the strain of the QWs. A larger hybridization gap leads to an increasing edge $v_F$. Electron-electron interaction effects in the helical edge can be parameterize by $K$, and $K$ is strongly correlated with $v_F$ and other factors such as screening from the environment [19,20]. In general, the helical Luttinger liquid has several fix points in the axis of $K$, namely, $K = 1$, $K = 1/2$, and $K = 1/4$. For helical edge states in (regular) InAs/GaSb QWs [24] assuming $v_F \sim 5 \times 10^4$ ms$^{-1}$, we have determined $K \sim 0.22$, which is close to 1/4.

As for the strained-layer InAs/GaInSb system, if we adopt a hybridization gap ~ 20 meV, $n_{cross}$ from (1 to 2) × 10$^{11}$ cm$^{-2}$, QWs width ~ 12 nm, and screening length ~ 50 nm, the estimated $v_F$ of helical edge states ranges then from (~ 1.9 to ~ 1.3) × 10$^5$ ms$^{-1}$, and consequently the $K$ value from ~ 0.5 to ~ 0.43. Thus even without further refining, the present system should cover the range of 1/4 (strongly-interacting) through 1/2 (weakly-interacting) in Luttinger parameter.

***Summary***   By strain-engineering, we have demonstrated for the first time a QSHI in InAs/GaInSb QWs clearly manifesting TRS protection, which shows a larger hybridization gap and longer characteristic length than existing QSHI systems. Moreover, the bulk is enough insulating at low temperatures and the edge characteristic length may be well controlled by the gates; data shows that the edge states can be gapped out by applying magnetic fields. Our findings not only move one step closer to the device and circuit applications of QSHI based on semiconductor technology, but also provide a nearly idea system for creating, detecting, and manipulating Majorana or parafermion bound states.

***Acknowledgments*** We thank Carlo W. J. Beenakker, Xincheng Xie, and Shoucheng Zhang for helpful discussions. Work at Rice were funded by NSF Grants No. DMR-1508644 and Welch Foundation Grants No. C-1682, work at PKU were funded by NBRPC Grants No. 2014CB920901, work at IoS, CAS were funded by NSFC Grants No. 11434010 and NBRPC Grants No. 2015CB921503.

**Figure Captions**

**Fig. 1. Calculated band dispersions and wafer structures of the strained InAs/GaInSb QWs.** (**a-c**) Calculated bulk band structure of the InAs/Ga$_{0.80}$In$_{0.20}$Sb (8.7 nm/4 nm) QWs, InAs/Ga$_{0.75}$In$_{0.25}$Sb (9 nm/4 nm) QWs, and InAs/Ga$_{0.68}$In$_{0.32}$Sb (8 nm/4 nm) QWs, CB1, VB1 and CB2, VB2 are bands of different spin component. (**d**) Schematic drawing of band dispersion (both bulk states and edge states) in InAs/GaInSb QSHI system. (**e**) Wafer structures of the strained-layer InAs/Ga$_{0.75}$In$_{0.25}$Sb QWs used for experiments. (**f**) Shown here as an example, a TEM photograph of the strained InAs/Ga$_{0.68}$In$_{0.32}$Sb wafer; blue and red lines are guide for eyes.

**Fig. 2. Transport data of bulk states from Corbino devices.** $G$-$V_{\text{front}}$ traces of a InAs/Ga$_{0.75}$In$_{0.25}$Sb Corbino under different in-plane magnetic field at (**a**) $V_{\text{back}}$ = 0 V and (**b**) $V_{\text{back}}$ = 4 V. $G$-$V_{\text{front}}$ traces under different perpendicular magnetic field at (**c**) $V_{\text{back}}$ = 0 V and (**d**) $V_{\text{back}}$ = 4 V. $G$-$V_{\text{front}}$ traces under different in-plane magnetic field for (**e**) a InAs/Ga$_{0.80}$In$_{0.20}$Sb Corbino and (**f**) a InAs/Ga$_{0.68}$In$_{0.32}$Sb Corbino. (**g**) Arrhenius plots for InAs/GaSb QWs (open squares), InAs/Ga$_{0.80}$In$_{0.20}$Sb QWs (open circles), InAs/Ga$_{0.75}$In$_{0.25}$Sb QWs (filled diamonds), and InAs/Ga$_{0.68}$In$_{0.32}$Sb QWs (filled circles). Energy gaps are deduced by fitting $G_{xx} \propto \exp(-\Delta/2k_BT)$, as shown by straight dash lines in the plot.

**Fig. 3. Helical edge transport in strained-layer InAs/Ga$_{0.75}$In$_{0.25}$Sb QWs.** $R_{xx}$-$V_{\text{front}}$ traces measured from (**a**) a 100 μm × 50 μm Hall bar device and (**b**) a 10 μm × 5 μm Hall bar device at $T$ ~ 20 mK with $V_{\text{back}}$ = 0 V, 1V, 2 V, 3 V, and 4 V. The edge characteristic length increases with decreasing $V_{\text{back}}$. The inset in (**a**) shows the $\lambda_\varphi$ and the $n_{\text{cross}}$ values at different backgate bias $V_{\text{back}}$.

**Fig. 4. Helical edge conductance under magnetic field.** (**a**) $R_{xx}$-$V_{\text{front}}$ traces of a 3 μm × 1.5 μm Hall bar under different perpendicular magnetic field ($V_{\text{back}}$ = 0 V). Inset in (**a**), the $R_{\text{CNP}}$ values increase monotonically in the range of $B_\perp$ < 5 T, manifesting enhanced backscattering processes in the helical edge under TRS breaking. (**b**) $R_{xx}$-$V_{\text{front}}$ traces of the 3 μm × 1.5 μm

Hall bar under different in-plane magnetic field. The plateau resistance values rise above the quantized value under $B_{//}$ because of TRS breaking. $R_{xx}$-$V_{front}$ traces of the 100 μm × 50 μm Hall bar under different perpendicular magnetic field at (**c**) $V_{back}$ = 0 V and (**d**) $V_{back}$ = 4 V.

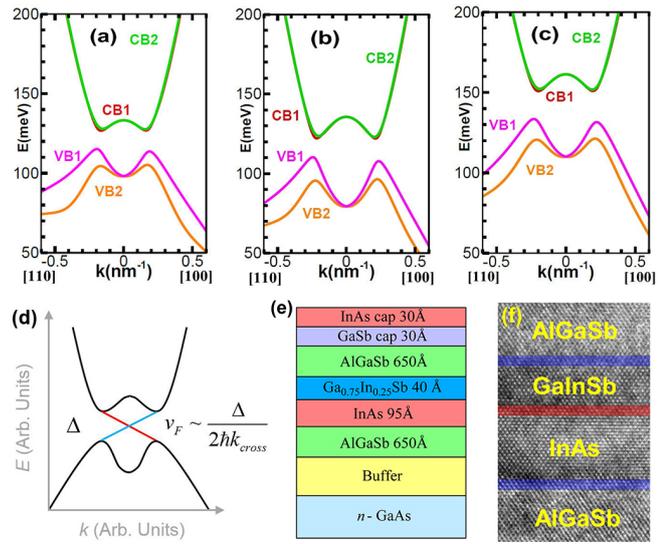

**Figure 1**

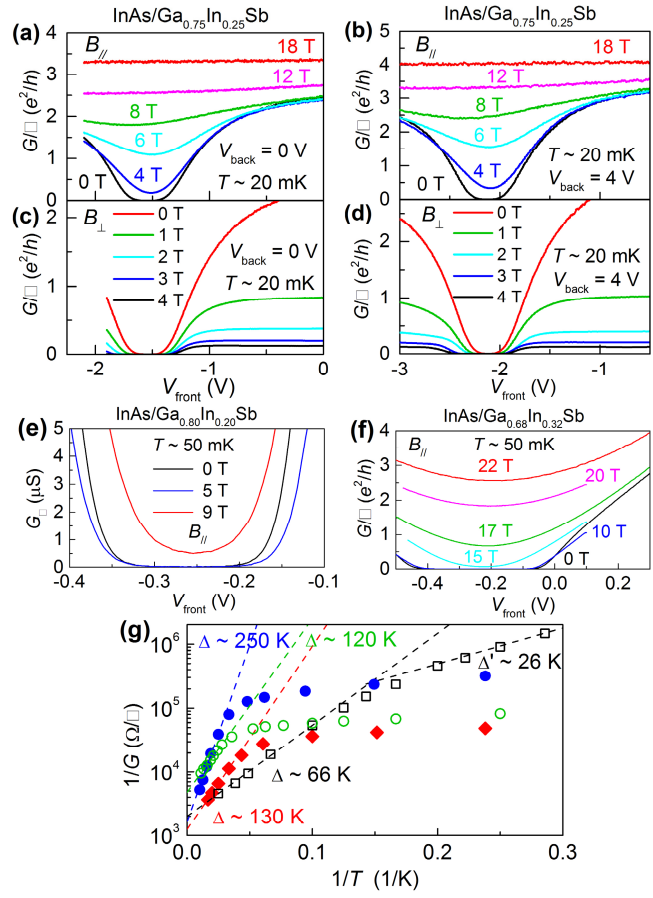

Figure 2

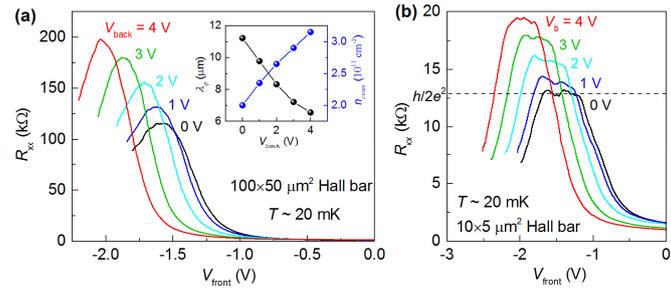

**Figure 3**

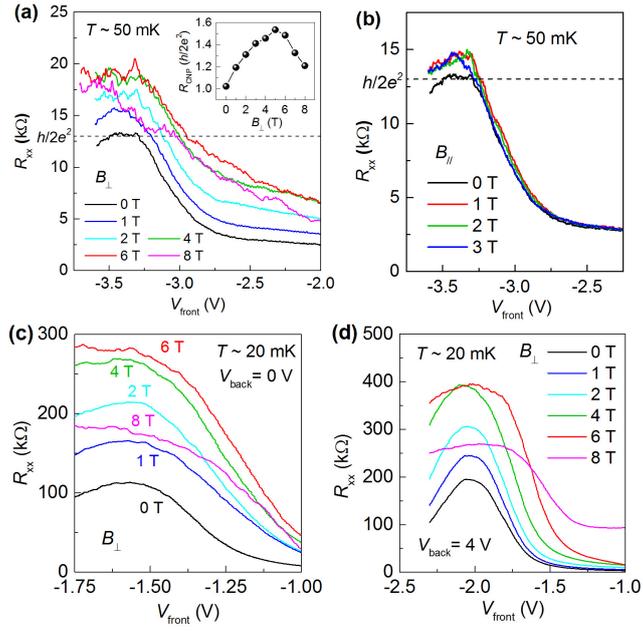

**Figure 4**

**Supplemental Material**

# Tuning Edge States in Strained-Layer InAs/GaInSb Quantum Spin Hall Insulators

Lingjie Du[1†], Tingxin Li[1, 2 †], Wenkai Lou[3], Xingjun Wu[2], Xiaoxue Liu[2], Zhongdong Han[2], Chi Zhang[2,4], Gerard Sullivan[5], Amal Ikhlassi[5], Kai Chang[3], Rui-Rui Du[1,2,4*]

† These authors contributed equally to this work.
* Correspondent author: Rui-Rui Du at rrd@rice.edu

[1]*Department of Physics and Astronomy, Rice University, Houston, Texas 77251-1892, USA*
[2]*International Center for Quantum Materials, School of Physics, Peking University, Beijing 100871, China*
[3]*SKLSM, Institute of Semiconductors, Chinese Academy of Sciences, Beijing 100083, China*
[4]*Collaborative Innovation Center of Quantum Matter, Beijing 100871, China*
[5]*Teledyne Scientific and Imaging, Thousand Oaks, California 91603, USA*

**I Wafer characterizations**

The semiconductor wafer of InAs/Ga$_{0.75}$In$_{0.25}$Sb QWs was grown by MBE technique. Fig. S1 shows the $R_{xx}$-$V_{front}$ trace and $B/eR_{xy}$-$V_{front}$ trace of a 30 μm × 10 μm Hall bar device measured at $T \sim 50$ mK, $V_b = 0$, and $B_\perp = 2$ T. It can be seen that there is a singularity of $B/eR_{xy}$, corresponding to $R_{xy} = 0$. Based on the classical two-carrier transport model, the Hall resistance $R_{xy}$ is given as:

$$R_{xy} = \frac{B[(p\mu_h^2 - n\mu_e^2) + \mu_e^2\mu_h^2 B^2(p-n)]}{e[(n\mu_e + p\mu_h)^2 + \mu_e^2\mu_h^2 B^2(p-n)^2]} \quad (1)$$

where $n$ and $p$ are electron and hole densities, and $\mu_e$ and $\mu_h$ are electron and hole mobilities, respectively; $B$ is the perpendicular magnetic field. Therefore, $R_{xy} = 0$ happens for the case $p > n$, since the $\mu_h$ is lower than the $\mu_e$, while $R_{xx}$ peak value appears when $p = n \neq 0$, thus the singularity of $R_{xy}$ should be on the left of the $R_{xx}$ peak, as shown in Fig. S1. On the other hand, if the bulk band is non-inverted, i.e. single carrier regime, both $R_{xy} = 0$ and $R_{xx}$ peak value should emerge for the case $p = n = 0$. In conclusion, data shown in Fig. S1 has confirmed that the bulk band structure of the present InAs/Ga$_{0.75}$In$_{0.25}$Sb QWs is inverted.

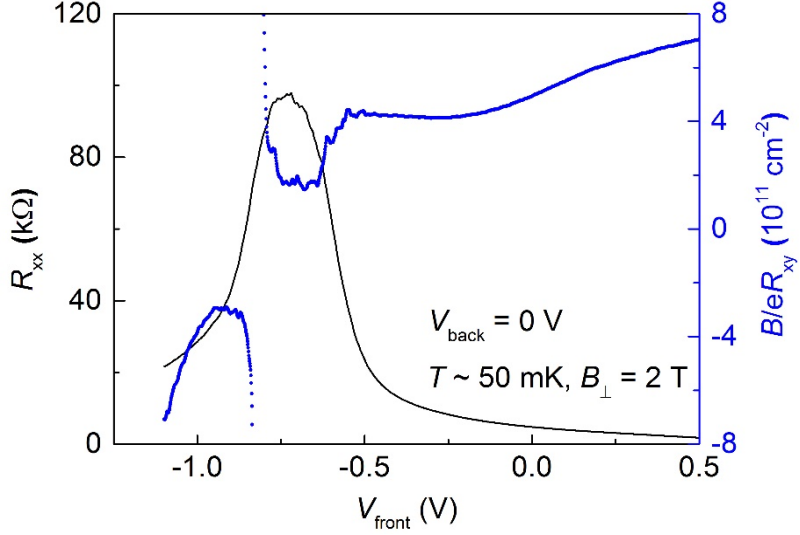

Fig. S1: $R_{xx}$-$V_{front}$ trace and $B/eR_{xy}$-$V_{front}$ trace of a 30 μm × 10 μm Hall bar.

We note that for those samples having the band structures in the semiconductor gap regime, the $R_{xy}$ should diverge near the edge of the gap, and consequently the $B/eR_{xy}$ approaches $0^+$ and $0^-$, respectively, as shown in ref. 6. Standard $R_{xy}$ measurement should provide unambiguous identification distinguishing between non-trivial gap in inverted band structure and trivial gap in semiconductor band.

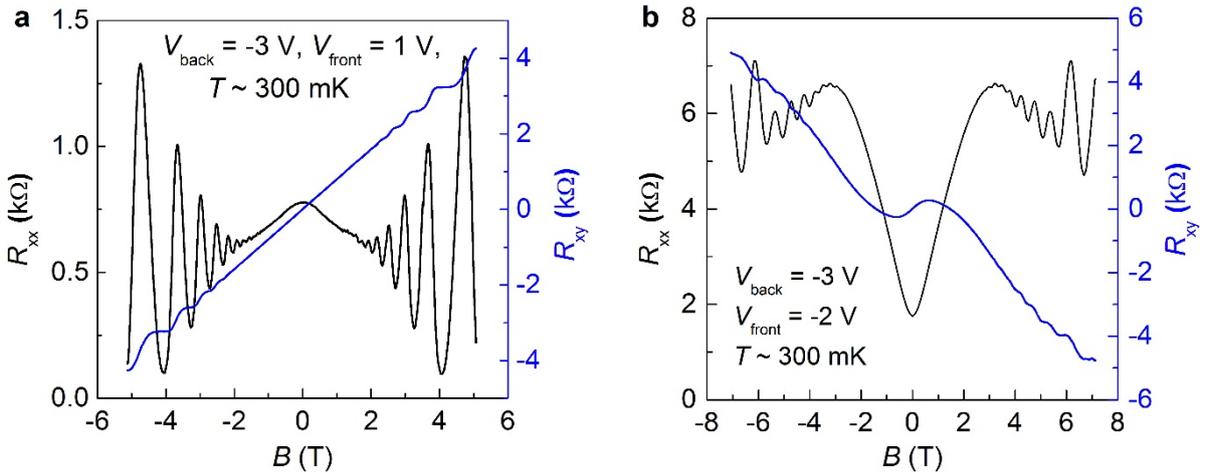

Fig. S2: Magneto-transport data of the 30 μm × 10 μm Hall bar in **a,** electron-dominant regime and **b,** hole-dominant regime.

Fig. S2a and S2b are two typical magneto-transport traces taken from the same device shown in Fig. S1. Fig. S2a is for the electron-dominant regime, where the electron density deduced from the SdH oscillations is ~ $7.8 \times 10^{11}$ cm$^{-2}$, with mobility ~ 30,000 cm$^2$/Vs. Fig. S2b is for the hole-dominant regime. In addition, the $R_{xy}$-$B$ trace in Fig. S2b clearly shows deviations from linear-dependence, indicating the two-carrier transport regime.

**II More data of Corbino devices**

Fig. S3 shows $G$-$V_{front}$ traces of a Corbino device at different temperatures (below 500 mK). Although a large hybridization gap is formed in the strained InAs/Ga$_{0.75}$In$_{0.25}$Sb QWs, bulk is only truly insulating at very low temperatures. At ~ 500 mK, the bulk resistance per square has already decreased to ~ 300 kΩ.

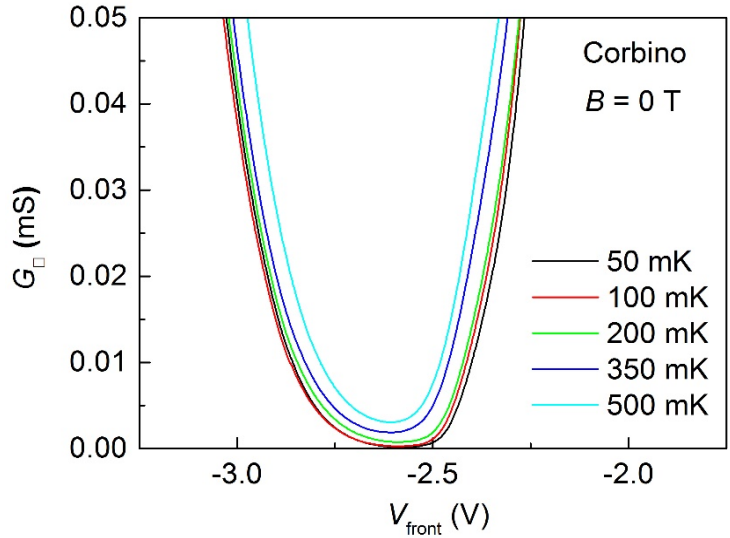

Fig. S3: $G$-$V_{front}$ traces of a Corbino device at 50 mK, 100 mK, 200 mK, 350 mK, and 500 mK.

**III More data of magnetic field dependence for Hall bar devices**

Fig. S4a and S4b show the $R_{xx}$-$V_{front}$ traces of the 100 μm × 50 μm Hall bar (mentioned in the main text) under in-plane magnetic field. It can be seen that the measured resistance peaks decrease, due to the bulk becoming conductive (semi-metallic) under $B_{//}$.

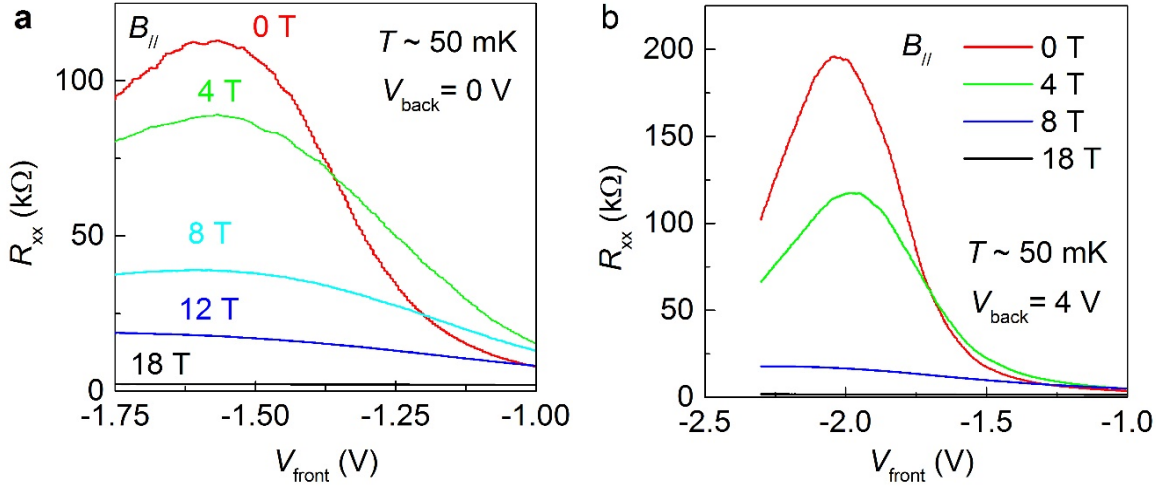

Fig. S4: $R_{xx}$-$V_{front}$ traces of the 100 μm × 50 μm Hall bar under different in-plane magnetic field at $V_{back}$ = 0 V (shown in **a**) and $V_{back}$ = 4 V (shown in **b**), respectively.

Fig. S5 shows the $R_{CNP}$ - $B_{//}$ trace measured from a 20 μm × 10 μm Schottky gated Hall bar device. We hold the $V_{front}$ at the $R_{xx}$ peak during the measurements. Clearly, the measured resistance increase at first due to TRS breaking, then decrease because of bulk conductance.

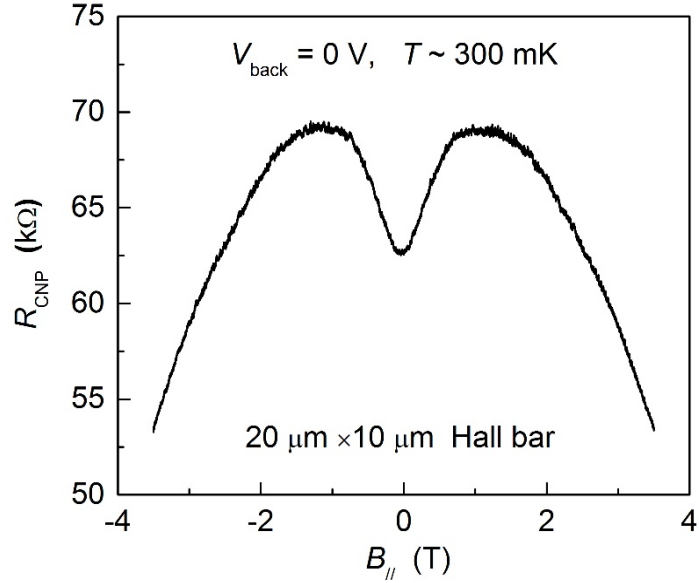

Fig. S5: $R_{CNP}$ - $B_{//}$ of a 20 μm × 10 μm Schottky gated Hall bar device, measured at 300 mK.

**V Estimations of Luttinger Parameters**

$K$ in a QSHI can be estimated by (*19,20*)

$$K = \left[1 + \frac{2}{\pi^2} \frac{e^2}{\varepsilon \hbar v_F} \ln\left(\frac{d}{\max\{\xi, w\}}\right)\right]^{-1/2} \quad (2)$$

where $\varepsilon$ is the bulk dielectric constant; $d$ is the distance from the QWs layers to a nearby metallic gate acts as a screening length for Coulomb potential; $w$ is the thickness of the QWs; assuming a linearly dispersing helical edge state, hence $v_F = \frac{1}{\hbar}\frac{\partial E}{\partial k} \sim \frac{E_{gap}}{2\hbar k_{cross}}$, where $v_F$ is the Fermi velocity of the helical edge state, $E_{gap}$ is the energy gap of the bulk QSHI, and $k_{cross} = \sqrt{2\pi n_{cross}}$; $\xi = 2\hbar v_F / E_{gap}$ is the evanescent decay length of the edge state wave function into the bulk QSHI.

For HgTe QWs (*19,20*),

$\varepsilon \approx 15$, $v_F \approx 5.5 \times 10^5$ m/s, $\xi \approx 30$ nm, $d \approx 150$ nm, $w \approx 12$ nm, so $K \approx 0.8$;

For shallow inverted InAs/GaSb QWs (*24*),

$\varepsilon \approx 12.5$, $v_F \approx 5.7 \times 10^4$ m/s, $\xi \approx 16$ nm, $d \approx 260$ nm, $w \approx 20$ nm, so $K \approx 0.22$.

For strained InAs/GaInSb QWs (if we adopt hybridization gap ~ 20 meV, screening length ~ 50 nm, QWs width ~ 12 nm, and $n_{cross}$ from $1 \times 10^{11}$ cm$^{-2}$ to $2 \times 10^{11}$ cm$^{-2}$),

$\varepsilon \approx 12.5$, $v_F \approx (1.9 - 1.3) \times 10^5$ m/s, $\xi \approx 12$ nm, $d \approx 50$ nm, $w \approx 12$ nm, so $K \approx 0.5 - 0.43$.